\documentstyle[11pt]{article} 
\begin{document} 
\begin{center} 
\hspace*{10mm} 

{\bf Chiral Symmetry Restoration through Hawking-Unruh Thermalization Effect}\\

\vspace{3mm} 

{Tadafumi Ohsaku}\\ 

\vspace{2mm} 
{\em Yukawa Institute for Theoretical Physics, Kyoto University, Kyoto, Japan, 
and Research Center for Nuclear Physics, Osaka University, Osaka, Japan}\\

\end{center} 

\vspace{3mm} 

The concepts of chiral, chiral symmetry and/or chiral rotation appear in a large part of 
theoretical physics~[1].
We can see a wide variety of usage in, for example, 
superstrings/M-theory ( holomorphic and antiholomorphic )~[2], 
chiral theories of grand unification~[3], supergravity ( superconformal tensor calculus~[4] ),
chiral symmtery breaking and its restoration in phenomena related to hadron and QCD~[5], 
nucleus and nuclear matter~[6,7,8], and also in superconductivity of condensed matter~[9]. 
Recently, I have published a paper on the dynamical chiral symmetry breaking and its restoration
for a uniformly accelerated observer~[10]. Here, I would like to make a short explanation
on it, and also give a speculation on the extension of its supersymmeric version.

The Hawking-Unruh effect predicts that 
an accelerated observer sees a thermal excitation of particles of a system in Minkowski spacetime,
with the Unruh temperature $T_{U}=a/2\pi$ ( $a$ is an acceleration constant )~[11].
This is called the thermalization theorem.
To study the effect of uniform acceleration, we employ Rindler metric embeded in flat Minkowski spacetime,
which are appropriate for the spacetime for a uniformly accelerated observer.
It should be noticed that the thermally excited particles are not the original Minkowski particles
but the Rindler particles determined 
by the ground ( vacuum ) state of the system of Rindler coordinates. 
The coordinate transformation arises the different picture for particle.
In fact, the structure of the spacetime which has a suitable boundary condition ( the event holizon ) is 
the origin of the thermal spectrum. 
It has been generally proved that, in Rindler coordinates, Euclidean two-point functions are 
periodic in the direction of time with period $2\pi$, and
these functions satisfy the Kubo-Martin-Schwinger condition.
Therefore, Green's functions in Rindler coordinates with period $2\pi$ 
may be interpreted as finite-temperature Green's functions.

To examine what will be observed by a uniformly accelerated observer
who moves relative to a system in which the chiral symmetry was dynamically broken,
the following situation is considered: After a fermion system in Minkowski spacetime 
dynamically generated a chiral mass, an observer will be accelerated uniformly.
The acceleration may give a thermal effect on the observation of the system.
Of particular interest here is the question whether the thermalization effect 
of acceleration gives the restoration of broken symmetry or not.
For the purpose to study the problem,
we employ the Nambu$-$Jona-Lasinio model.
The field theory in Rindler coordinates can be described by the method
of curved background field. 
We obtain the Green's function in Rindler coordinates, 
and apply the method of effective potential for our problem.
The results are summarized as follows: 
(1) The gap equation for a self-consistent mass are obtaind in Rindler coordinates. 
(2) The critical coupling and the critical acceleration ( the critical Hawking-Unruh temperature ) 
for the symmetry breaking and its restoration are obtained. The results coincide with
the case of usual temperature restoration in Minkowski spacetime. 
(3) The critical exponent shows simply the case of the Landau theory of second order
phase transition.

On the other hand, recently a paper on the phase transition from color glass condensate 
to quark gluon plasma appeared by Karzeev and Tuchin~[12]. 
In that paper, the possible scenario of thermalization
from the Hawking-Unruh effect in relativistic heavy ion collision was proposed. 
They argued that the transition temperature
through the mechanism of the Hawking-Unruh thermalization
can become an experimentally obtainable value under suitable situations of 
relativistic heavy ion collisions.

Now, it is the stage for us to make final remarks.
The investigation of the chiral restoration of the Hawking-Unruh effct in 
the supersymmetric Nambu$-$Jona-Lasino ( SNJL ) model can easily be done~[13].
Without a lengthy calculation, we can speculate the results.
In the SNJL model, the dynamical chiral symmetry breaking can only occur
under the supersymmetry is broken. Thus we should employ the SNJL model
with soft SUSY mass. With the same method explained above, 
the gap equation, the critical coupling and the critical acceleration
could be obtained. In this case, the relation between the coupling, the cutoff and the strength
of soft mass is important to realize the dynamical chiral symmetry breaking.
If the soft mass is large enough, the chiral symmetry will be dynamically broken.
To consider the situations of the order of the phase transition,
only we need the spectrum of thermal excitation. 
In this case, the order of the phase transition under Hawking-Unruh thermalization
changes from first-order to second-order when the soft mass becomes large enough.
With taking into account the result of the paper of Karzeev and Tuchin, 
the study of the phase transition from color grass condensates to quark gluon plasma by 
SNJL will be an important subject for us. 
The quantitative study for this problem is now under preparation.

\end{document}